\documentclass[aip,rsi,reprint]{revtex4-1} 

\usepackage{siunitx}
\usepackage{amsmath}
\usepackage{amsfonts}
\usepackage{txfonts}
\usepackage{graphicx}
\usepackage[caption=false]{subfig}
\usepackage{dcolumn}
\usepackage{multirow}
\usepackage{hyperref}

\hypersetup{
    colorlinks=true,       
    linkcolor=blue,          
    citecolor=blue,        
    filecolor=blue,      
    urlcolor=blue           
}

\DeclareGraphicsExtensions{.png,.tiff,.ai,.eps}

\graphicspath{{./}{./Figures/}}

\usepackage{natbib}

\begin{document}

\title{A radiofrequency voltage-controlled current source for quantum spin manipulation}

\author{D.S. Barker}
\email[]{daniel.barker@nist.gov}
\author{A. Restelli}
\email[]{arestell@umd.edu}
\affiliation{Joint Quantum Institute, University of Maryland and National Institute of Standards and Technology \\ College Park, MD 20742, USA}
\author{J.A. Fedchak}
\author{J. Scherschligt}
\author{S. Eckel}
\affiliation{Sensor Science Division, National Institute of Standards and Technology \\ Gaithersburg, MD 20899, USA}

\date{\today}

\begin{abstract}

We present a wide-bandwidth, voltage-controlled current source that is easily integrated with radiofrequency magnetic field coils.
Our design uses current feedback to compensate for the frequency-dependent impedance of a radiofrequency antenna.
We are able to deliver peak currents greater than \(100~\si{\milli\ampere}\) over a \(300~\si{\kilo\hertz}\) to \(54~\si{\mega\hertz}\) frequency span.
The radiofrequency current source fits onto a printed circuit board smaller than \(4~\si{\square\centi\meter}\) and consumes less than \(1.3~\si{\watt}\) of power.
It is suitable for use in deployable quantum sensors and nuclear-magnetic-resonance systems.

\end{abstract}

\maketitle

\section{Introduction}

Radiofrequency (RF) manipulation of electronic and nuclear spins is essential to many fields.
Nuclear magnetic resonance (NMR) requires RF excitation to initialize spin precession and has applications including magnetic resonance imaging,~\cite{Plewes2012, Grover2015} chemical analysis,~\cite{Lee2008} inertial navigation,~\cite{Kitching2018} and magnetometry.~\cite{Kitching2018}
Radiofrequency magnetic fields are also an important tool in atomic physics research, where they are used to prepare~\cite{Myatt1997, Stan2005, Greif2007, Brown2015, Geiger2018} and control~\cite{Krauser2014, Lundblad2014, Anderson2018, Trypogeorgos2018} quantum gases.
Broadband RF excitation is necessary for many applications in NMR and atomic physics.
For example, multinuclear NMR requires RF tones spaced by approximately \( 10~\si{\mega\hertz}\) per tesla of bias field~\cite{Davoodi2019} and RF-induced evaporative cooling typically requires that the excitation frequency be swept over a range wider than \(20~\si{\mega\hertz}\).~\cite{Myatt1997, Stan2005, Greif2007, Brown2015}

Experimentalists typically use RF power amplifiers connected to small loop or transmission line antennas to create RF magnetic fields.~\cite{Myatt1997, Stan2005, Greif2007, Murphree2007, Brown2015, Geiger2018}
To prevent damage to the RF amplifier, the antenna's impedance must be matched to \(50~\si{\ohm}\) to minimize power reflection.
Loop and transmission line antennas can both be impedance matched over a reasonably wide frequency bandwidth by terminating one port of the antenna with a \(50~\si{\ohm}\) resistor.~\cite{Murphree2007, Anders2018, Davoodi2019}
Resistive impedance matching maximizes the power transmitted through the antenna, but limits the transmitted current.
Because the RF magnetic field amplitude depends on the current, resistively matched antennas typically require large, high-power amplifiers to produce the necessary RF field strength.~\cite{Stan2005, Greif2007, Brown2015}
As atomic physics and NMR systems move from laboratories into the field, the size, weight, and power dissipation (SWaP) of RF-field generating electronics will need to be improved.

One approach to reducing the size of RF amplification systems is co-locating a RF metal-oxide-semiconductor field-effect transistor (MOSFET) and the antenna.~\cite{Copeland1964, Kurpad2006, Heilman2007}
Integration of the amplifier with the antenna allows the RF MOSFET to act as a voltage-controlled current source (VCCS).~\cite{Kurpad2006, Heilman2007, Lee2009}
The power dissipation for the desired drive current can also be reduced in RF VCCS designs using class AB or class D amplifier layouts.~\cite{Heilman2007, Lee2009, Gudino2013}
Radiofrequency voltage-controlled current sources are often used for magnetic resonance imaging (MRI) because, in addition to better SWaP, directly controlling the antenna current reduces cross-talk; leading to improved magnetic field uniformity.~\cite{Kurpad2006, Lee2009, Gudino2013}
To ensure that the RF MOSFET behaves as a RF current source, the antenna's input impedance must be small compared to the MOSFET's output impedance.
Because both the MOSFET and antenna impedances have significant reactive components, MOSFET-based RF voltage-controlled current sources have a narrow operating bandwidth and are not suitable for broadband NMR and atomic physics applications.

We realize a wide-bandwidth, voltage-controlled RF current source.
Our design employs current feedback to compensate for the frequency-dependent impedance of the RF antenna.
It can drive peak currents \(I_p > 100~\si{\milli\ampere}\) through a \(300~\si{\nano\henry}\) test antenna from \(300~\si{\kilo\hertz}\) to \(54~\si{\mega\hertz}\).
The VCCS fits onto a printed circuit board (PCB) smaller than \(4~\si{\square\centi\meter}\), making it compatible with compact quantum devices and NMR systems.
The design files for the RF VCCS, including schematics, board layouts, and bills of materials, are available online.~\cite{vccsgit}

\section{Design}

The active element in our design is a current-feedback operational amplifier (CFA).
In contrast to the more common voltage-feedback operational amplifier (VFA), a CFA buffers the voltage applied to its noninverting input onto its inverting input.~\cite{Karki1998, Mancini2003}
The CFA then attempts to drive the buffer's output current \(I_e\) to zero by outputing a voltage \(V_{\rm out}=I_e Z_t(\omega)\), where \(Z_t(\omega)\) is the transimpedance of the CFA and \(\omega\) is the angular frequency. 
The current-feedback architecture has two advantages that make CFAs ideal for a RF VCCS.\@
First, CFAs can achieve larger slew rates than VFAs, so they can drive large signals at higher frequencies.
Second, the maximum closed-loop bandwidth of a CFA is controlled by the feedback impedance \(Z_f\), rather than a fixed gain-bandwidth product.
The independence of gain and bandwidth for CFAs allows us to match the input dynamic range of the VCCS to the output dynamic range of a RF signal source, such as a direct digital synthesizer.

\begin{figure}[t]
\includegraphics[width=\linewidth]{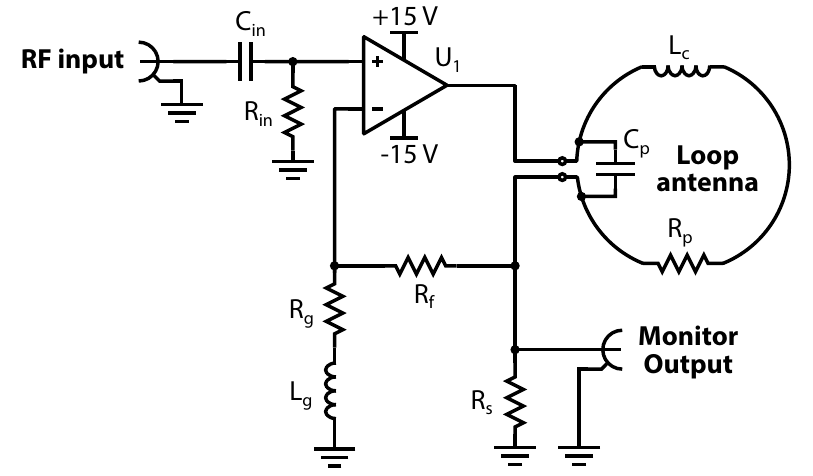}
\caption{\label{Fig:Circuit}
Circuit diagram of the RF voltage-controlled current source.
}
\end{figure}

Figure~\ref{Fig:Circuit} shows the circuit diagram of our voltage-controlled current source.
We layout the circuit so the THS3091 current-feedback operational amplifier \(U_1\) acts as a non-inverting amplifier.~\cite{ths3091, disclaimer}
Feedback and gain-setting resistors (\(R_f=215~\si{\ohm}\) and \(R_g=23.7~\si{\ohm}\), respectively) set the non-inverting DC gain \(G=1+R_f/R_g\approx 10\).
The input resistor and capacitor (\(R_{\rm in}\) and \(C_{\rm in}\)) match the non-inverting input impedance of the CFA to \(50~\si{\ohm}\) and high-pass filter the RF voltage-control input.
We represent our loop antenna as an inductor \(L_c\) with series resistance \(R_p\) and parasitic capacitance \(C_p\).
The resistor \(R_s\) senses the current driven through the antenna by \(U_1\).
We choose \(R_f\), \(R_g\), and \(R_s=10~\si{\ohm}\) so that the DC transadmittance of the VCCS is \(Y(\omega = 0)=G/R_s\approx 1~\si[per-mode=symbol]{\ampere\per\volt}\).

\begin{figure}[t]
\includegraphics[width=\linewidth]{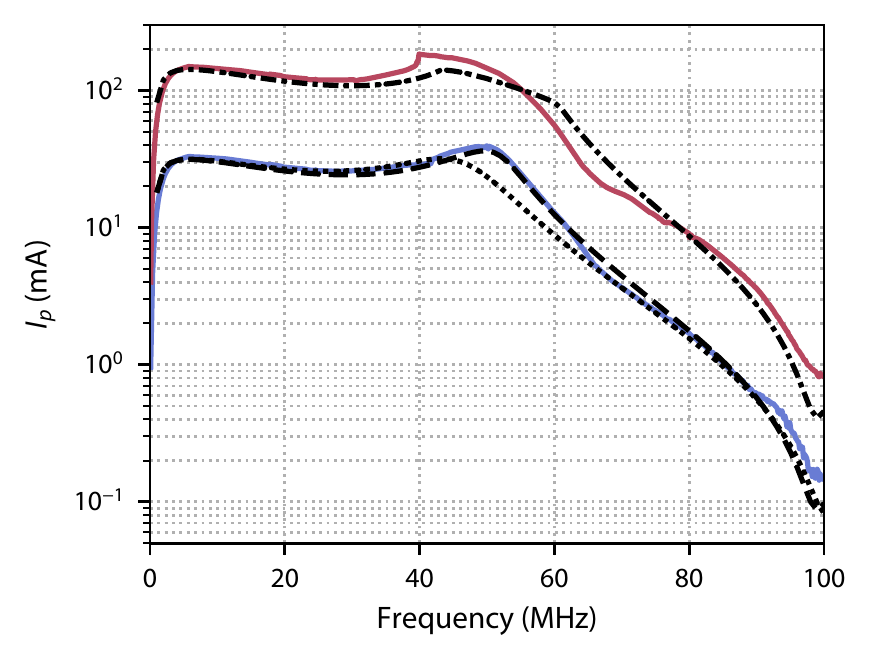}
\caption{\label{Fig:CurrentvFreq}
Frequency response of the RF voltage-controlled current source.
The blue (red) curve shows the amplitude of the current \(I_p\) passing through the test antenna for a RF input voltage amplitude \(V_p = 28~\si{\milli\volt}~(127~\si{\milli\volt})\).
The nominal current amplitude \(I_p\) for the blue and red curves is \(34~\si{\milli\ampere}\) and \(152~\si{\milli\ampere}\), respectively.
Both curves exhibit approximately \(\pm 2~\si{decibel}\) of ripple in the passband.
The dashed (dash-dotted) curves are the result of lumped-element transient simulations of the small-signal (large-signal) response of the circuit in Fig.~\ref{Fig:Circuit} using the populated component values.
The dotted curve shows the response expected from the transadmittance \(Y(\omega)\) derived using the circuit's closed-loop transfer function.
}
\end{figure}

The prescence of the loop antenna within the CFA's feedback loop causes the current source's AC response to deviate from that of a simple non-inverting amplifier.
We model the CFA's transimpedance as \(Z_t(\omega) = R_t/(1+j \omega R_t C_t)\), where \(j\) is the imaginary unit, \(R_t \approx 850~\si{\kilo\ohm}\) is the amplifier's DC transimpedance, and the capacitance \(C_t \approx 1~\si{\pico\farad}\) sets the amplifier's open-loop bandwidth.~\cite{Karki1998, ths3091}
The loop inductance \(L_c\) and transimpedance \(Z_t(\omega)\) form a \(RLC\) circuit that adds a pair of conjugate poles to the transadmittance \(Y(\omega)\).
The poles are underdamped, so they create a sharp peak in \(Y(\omega)\) near the natural frequency
\begin{equation}
  \omega_n = \sqrt{\frac{1}{L_c C_t}}\sqrt{\frac{1+\frac{R_f}{R_t}+\frac{R_i}{R_g}\frac{R_f+R_g}{R_t}}{1+\frac{R_f}{R_s}+\frac{R_i}{R_g}\frac{R_s+R_f+R_g}{R_s}}},
\end{equation}
where \(R_i \approx 30~\si{\ohm}\) is the inverting input impedance of the CFA and we have assumed that \(R_p\) and \(C_p\) are negligible.~\cite{ths3091}
To flatten the response of the current source, we use a gain compensation inductor \(L_g\) in series with \(R_g\) to introduce a real pole to the RF VCCS transfer function.
The natural frequency of the new pole is approximately \(R_g/L_g\), so reduction of the transadmittance peak near \(\omega_n\) requires \(L_g \ge R_g/\omega_n\).
Larger values of \(L_g\) give greater suppression of the peak in \(Y(\omega)\), but also increase undershoot.
By varying populated and simulated component values, we find empirically that \(L_g \approx 2 R_g/\omega_n\) is a reasonable compromise between overshoot and undershoot (\(L_g = 2.2 R_g/\omega_n\) for the data in Fig.~\ref{Fig:CurrentvFreq}).

The contribution of \(L_c\) to the feedback impedance \(Z_f\) also constricts the bandwidth of the RF VCCS.\@
To alleviate the bandwidth constriction imposed by \(L_c\), we select \(R_f\) below the recommended value in the THS3091 datasheet.~\cite{ths3091}
Small values of \(R_f\) would normally cause the CFA to become unstable,~\cite{Karki1998, Mancini2003} but in our circuit \(Z_f \approx R_f + R_p + j\omega L_c\) increases to match the recommended value of \(R_f\) when \(\omega\) approaches the maximum bandwidth of the CFA.\@
The wide operating bandwidth of the RF VCCS is possible because the frequency-dependent impedance of \(L_c\) stabilizes the CFA.\@
The gain and phase margins for the RF VCCS are approximately \(30~\si{\decibel}\) and \(20\si{\degree}\), respectively.

\section{Results}

Figure~\ref{Fig:CurrentvFreq} shows current output of the RF VCCS as measured at the monitor output (see Fig.~\ref{Fig:Circuit}).
We plot the small-signal (blue) and large-signal (red) response versus the frequency of the RF input voltage.
The amplitude of the small-signal (large-signal) RF input voltage is \(V_p = 28~\si{\milli\volt}~(127~\si{\milli\volt})\), which corresponds to a commanded current amplitude \(I_p = 34~\si{\milli\ampere}~(152~\si{\milli\ampere})\).
The DC transadmittance increases to \(Y(\omega = 0)\approx 1.2~\si[per-mode=symbol]{\ampere\per\volt}\) because the \(50~\si{\ohm}\) impedance of our spectrum analyzer is in parallel with \(R_s\).
A DC block protecting the spectrum analyzer rolls-off the data at low frequency.
The low-frequency and high-frequency \(3~\si{\decibel}\) corners in the transadmittance are approximately \(300~\si{\kilo\hertz}\) and \(54~\si{\mega\hertz}\), respectively.
The transadmittance has approximately \(\pm 2~\si{\decibel}\) of ripple in the passband.

The dashed and dash-dotted curves in Fig.~\ref{Fig:CurrentvFreq} are the result of lumped-element transient simulations of the RF VCCS in the small- and large-signal limits, respectively.
We use the populated component values and the Texas Instruments SPICE model to represent the THS3091.~\cite{pspice}
The parasitic resistance \(R_p\) of the antenna's PCB trace is approximately \(100~\si{\milli\ohm}\).
We adjust \(L_c\) and \(C_p\) in the small-signal simulation to reproduce the positions of the local maximum (near \(50~\si{\mega\hertz}\)) and minimum (near \(100~\si{\mega\hertz}\)) in the small-signal transadmittance, respectively.
The small-signal simulation reasonably matches the data when we take \(L_c = 270~\si{\nano\henry}\) and \(C_p = 9~\si{\pico\farad}\).
(The SPICE model for the THS3091 does not accurately reproduce harmonic distortion leading to greater disagreement between our data and the large-signal simulation.~\cite{pspice})
The inductance agrees reasonably with the expected value \(L_c = 220~\si{\nano\henry}\) for a loop antenna with same radius and cross-sectional area.~\cite{Balanis2016}
We presume that the difference between the fitted and calculated values for \(L_c\) results from a combination of skin effects and stray inductance.

The performance of the RF VCCS can also calculated from the full transfer function for the circuit in Fig.~\ref{Fig:Circuit} (see supplementary material).
We use the closed-loop transfer function to determine the transadmittance \(Y(\omega)\).
The dotted curve in Fig.~\ref{Fig:CurrentvFreq} shows the small-signal response derived from \(Y(\omega)\), with \(L_c = 270~\si{\nano\henry}\) and \(C_p = 9~\si{\pico\farad}\).
The transfer function accurately predicts the behavior of the RF VCCS at low frequency, but underestimates the achievable bandwidth.

The test antenna limits both the bandwidth and the current output of the RF VCCS.~\footnote{The size of our antenna is set by geometric constraints of our laser-cooling apparatus.}
The impedance of the antenna approaches \(100~\si{\ohm}\) at \(50~\si{\mega\hertz}\), so attempting to drive currents larger than \(150~\si{\milli\ampere}\) reduces the large-signal bandwidth due to the limited output voltage swing of the CFA.\@
The antenna's high impedance also causes the feature in the large-signal response near \(40~\si{\mega\hertz}\) (see Fig.~\ref{Fig:CurrentvFreq}), where the voltage required to drive the commanded current exceeds the output voltage swing of the CFA.\@
At the feature's frequency, the RF VCCS enters a regime of non-linear operation and drives sub-harmonic oscillations.
The sub-harmonics only occur within a narrow band around the feature and can be eliminated by reducing the commanded current.
We have simulated the performance of the current source with smaller antennas (e.g.\ the microcoil design of Fratila et al.~\cite{Fratila2014}) and found that the RF VCCS bandwidth can approach the \(180~\si{\mega\hertz}\) bandwidth of the THS3091 (see supplementary material).
Using a smaller loop would also allow us to exploit the full \(250~\si{\milli\ampere}\) output capability of the THS3091.
Wider bandwidth and higher current output could be acheived by using a different CFA (e.g. THS3491~\cite{ths3491}).
We could also increase the output current by driving an antenna with several parallel CFAs, as suggested by the load-sharing circuit in the THS3091 datasheet.~\cite{ths3091}

\section{Conclusion}

We have demonstrated a wideband RF voltage-controlled current source with an integrated antenna.
The RF VCCS uses a CFA to deliver peak currents greater than \(100~\si{\milli\ampere}\) over a frequency range of \(300~\si{\kilo\hertz}\) to \(54~\si{\mega\hertz}\) with power consumption less than \(1.3~\si{\watt}\).
It is suitable for manipulating spin systems in the low magnetic fields typically used in quantum sensors.~\cite{Kitching2018, Eckel2018}
By using a more advanced coil design, the bandwidth of the RF current source could be increased, allowing its use in deployable NMR systems.
The design files for our RF VCCS are available online for others to use and modify.~\cite{vccsgit}

\section*{Supplementary Material}

The supplementary material contains derivations of the transfer functions of the RF VCCS and shows the simulated performance of the RF VCCS with lower inductance antennas.

\section*{Acknowledgements}

The authors thank K. Douglass and W. McGehee for their careful reading of the manuscript.

\section*{Data Availability}

The data that support our findings are openly available online.~\cite{vccsgit, doi}

\bibliography{RFVI}

\begin{thebibliography}{33}%
\makeatletter
\providecommand \@ifxundefined [1]{%
 \@ifx{#1\undefined}
}%
\providecommand \@ifnum [1]{%
 \ifnum #1\expandafter \@firstoftwo
 \else \expandafter \@secondoftwo
 \fi
}%
\providecommand \@ifx [1]{%
 \ifx #1\expandafter \@firstoftwo
 \else \expandafter \@secondoftwo
 \fi
}%
\providecommand \natexlab [1]{#1}%
\providecommand \enquote  [1]{``#1''}%
\providecommand \bibnamefont  [1]{#1}%
\providecommand \bibfnamefont [1]{#1}%
\providecommand \citenamefont [1]{#1}%
\providecommand \href@noop [0]{\@secondoftwo}%
\providecommand \href [0]{\begingroup \@sanitize@url \@href}%
\providecommand \@href[1]{\@@startlink{#1}\@@href}%
\providecommand \@@href[1]{\endgroup#1\@@endlink}%
\providecommand \@sanitize@url [0]{\catcode `\\12\catcode `\$12\catcode
  `\&12\catcode `\#12\catcode `\^12\catcode `\_12\catcode `\%12\relax}%
\providecommand \@@startlink[1]{}%
\providecommand \@@endlink[0]{}%
\providecommand \url  [0]{\begingroup\@sanitize@url \@url }%
\providecommand \@url [1]{\endgroup\@href {#1}{\urlprefix }}%
\providecommand \urlprefix  [0]{URL }%
\providecommand \Eprint [0]{\href }%
\providecommand \doibase [0]{http://dx.doi.org/}%
\providecommand \selectlanguage [0]{\@gobble}%
\providecommand \bibinfo  [0]{\@secondoftwo}%
\providecommand \bibfield  [0]{\@secondoftwo}%
\providecommand \translation [1]{[#1]}%
\providecommand \BibitemOpen [0]{}%
\providecommand \bibitemStop [0]{}%
\providecommand \bibitemNoStop [0]{.\EOS\space}%
\providecommand \EOS [0]{\spacefactor3000\relax}%
\providecommand \BibitemShut  [1]{\csname bibitem#1\endcsname}%
\let\auto@bib@innerbib\@empty
\bibitem [{\citenamefont {Plewes}\ and\ \citenamefont
  {Kucharczyk}(2012)}]{Plewes2012}%
  \BibitemOpen
  \bibfield  {author} {\bibinfo {author} {\bibfnamefont {D.~B.}\ \bibnamefont
  {Plewes}}\ and\ \bibinfo {author} {\bibfnamefont {W.}~\bibnamefont
  {Kucharczyk}},\ }\href {\doibase 10.1002/jmri.23642} {\bibfield  {journal}
  {\bibinfo  {journal} {Journal of Magnetic Resonance Imaging}\ }\textbf
  {\bibinfo {volume} {35}},\ \bibinfo {pages} {1038} (\bibinfo {year}
  {2012})}\BibitemShut {NoStop}%
\bibitem [{\citenamefont {Grover}\ \emph {et~al.}(2015)\citenamefont {Grover},
  \citenamefont {Tognarelli}, \citenamefont {Crossey}, \citenamefont {Cox},
  \citenamefont {Taylor-Robinson},\ and\ \citenamefont {McPhail}}]{Grover2015}%
  \BibitemOpen
  \bibfield  {author} {\bibinfo {author} {\bibfnamefont {V.~P.~B.}\
  \bibnamefont {Grover}}, \bibinfo {author} {\bibfnamefont {J.~M.}\
  \bibnamefont {Tognarelli}}, \bibinfo {author} {\bibfnamefont {M.~M.~E.}\
  \bibnamefont {Crossey}}, \bibinfo {author} {\bibfnamefont {I.~J.}\
  \bibnamefont {Cox}}, \bibinfo {author} {\bibfnamefont {S.~D.}\ \bibnamefont
  {Taylor-Robinson}}, \ and\ \bibinfo {author} {\bibfnamefont {M.~J.~W.}\
  \bibnamefont {McPhail}},\ }\href {\doibase 10.1016/j.jceh.2015.08.001}
  {\bibfield  {journal} {\bibinfo  {journal} {Journal of Clinical and
  Experimental Hepatology}\ }\textbf {\bibinfo {volume} {5}},\ \bibinfo {pages}
  {246} (\bibinfo {year} {2015})}\BibitemShut {NoStop}%
\bibitem [{\citenamefont {Lee}\ \emph {et~al.}(2008)\citenamefont {Lee},
  \citenamefont {Sun}, \citenamefont {Ham},\ and\ \citenamefont
  {Weissleder}}]{Lee2008}%
  \BibitemOpen
  \bibfield  {author} {\bibinfo {author} {\bibfnamefont {H.}~\bibnamefont
  {Lee}}, \bibinfo {author} {\bibfnamefont {E.}~\bibnamefont {Sun}}, \bibinfo
  {author} {\bibfnamefont {D.}~\bibnamefont {Ham}}, \ and\ \bibinfo {author}
  {\bibfnamefont {R.}~\bibnamefont {Weissleder}},\ }\href {\doibase
  10.1038/nm.1711} {\bibfield  {journal} {\bibinfo  {journal} {Nature
  Medicine}\ }\textbf {\bibinfo {volume} {14}},\ \bibinfo {pages} {869}
  (\bibinfo {year} {2008})}\BibitemShut {NoStop}%
\bibitem [{\citenamefont {Kitching}(2018)}]{Kitching2018}%
  \BibitemOpen
  \bibfield  {author} {\bibinfo {author} {\bibfnamefont {J.}~\bibnamefont
  {Kitching}},\ }\href {\doibase 10.1063/1.5026238} {\bibfield  {journal}
  {\bibinfo  {journal} {Applied Physics Reviews}\ }\textbf {\bibinfo {volume}
  {5}},\ \bibinfo {pages} {031302} (\bibinfo {year} {2018})}\BibitemShut
  {NoStop}%
\bibitem [{\citenamefont {Myatt}(1997)}]{Myatt1997}%
  \BibitemOpen
  \bibfield  {author} {\bibinfo {author} {\bibfnamefont {C.~J.}\ \bibnamefont
  {Myatt}},\ }\emph {\bibinfo {title} {{Bose-Einstein Condensation Experiments
  in a Dilute Vapor of Rubidium}}},\ \href@noop {} {Ph.D. thesis},\ \bibinfo
  {school} {University of Colorado} (\bibinfo {year} {1997})\BibitemShut
  {NoStop}%
\bibitem [{\citenamefont {Stan}(2005)}]{Stan2005}%
  \BibitemOpen
  \bibfield  {author} {\bibinfo {author} {\bibfnamefont {C.~A.}\ \bibnamefont
  {Stan}},\ }\emph {\bibinfo {title} {{Experiments with Interacting Bose and
  Fermi Gases}}},\ \href@noop {} {Ph.D. thesis},\ \bibinfo  {school}
  {Masshachusetts Institute of Technology} (\bibinfo {year} {2005})\BibitemShut
  {NoStop}%
\bibitem [{\citenamefont {Greif}(2007)}]{Greif2007}%
  \BibitemOpen
  \bibfield  {author} {\bibinfo {author} {\bibfnamefont {D.~G.}\ \bibnamefont
  {Greif}},\ }\emph {\bibinfo {title} {{Evaporative cooling and Bose-Einstein
  Condensation of Rb-87 in a moving-coil TOP trap geometry}}},\ \href
  {http://ultracold.physics.sunysb.edu/subpages/pubfiles/Greif{\_}MA2007.pdf}
  {Master's thesis},\ \bibinfo  {school} {Stony Brook University} (\bibinfo
  {year} {2007})\BibitemShut {NoStop}%
\bibitem [{\citenamefont {Brown}(2015)}]{Brown2015}%
  \BibitemOpen
  \bibfield  {author} {\bibinfo {author} {\bibfnamefont {R.~C.}\ \bibnamefont
  {Brown}},\ }\emph {\bibinfo {title} {{Nonequilibrium Manybody Dynamics with
  Ultracold Atoms in Optical Lattices and Selected Problems in Atomic
  Physics}}},\ \href@noop {} {Ph.D. thesis},\ \bibinfo  {school} {University of
  Maryland, College Park} (\bibinfo {year} {2015})\BibitemShut {NoStop}%
\bibitem [{\citenamefont {Geiger}(2018)}]{Geiger2018}%
  \BibitemOpen
  \bibfield  {author} {\bibinfo {author} {\bibfnamefont {Z.~A.}\ \bibnamefont
  {Geiger}},\ }\emph {\bibinfo {title} {{An Apparatus for Dynamical Quantum
  Emulation Using Ultracold Lithium}}},\ \href
  {http://web.physics.ucsb.edu/{~}weld/publications/Geiger.pdf} {Ph.D.
  thesis},\ \bibinfo  {school} {University of California, Santa Barbara}
  (\bibinfo {year} {2018})\BibitemShut {NoStop}%
\bibitem [{\citenamefont {Krauser}\ \emph {et~al.}(2014)\citenamefont
  {Krauser}, \citenamefont {Ebling}, \citenamefont {Fl{\"{a}}schner},
  \citenamefont {Heinze}, \citenamefont {Sengstock}, \citenamefont
  {Lewenstein}, \citenamefont {Eckardt},\ and\ \citenamefont
  {Becker}}]{Krauser2014}%
  \BibitemOpen
  \bibfield  {author} {\bibinfo {author} {\bibfnamefont {J.~S.}\ \bibnamefont
  {Krauser}}, \bibinfo {author} {\bibfnamefont {U.}~\bibnamefont {Ebling}},
  \bibinfo {author} {\bibfnamefont {N.}~\bibnamefont {Fl{\"{a}}schner}},
  \bibinfo {author} {\bibfnamefont {J.}~\bibnamefont {Heinze}}, \bibinfo
  {author} {\bibfnamefont {K.}~\bibnamefont {Sengstock}}, \bibinfo {author}
  {\bibfnamefont {M.}~\bibnamefont {Lewenstein}}, \bibinfo {author}
  {\bibfnamefont {A.}~\bibnamefont {Eckardt}}, \ and\ \bibinfo {author}
  {\bibfnamefont {C.}~\bibnamefont {Becker}},\ }\href {\doibase
  10.1126/science.1244059} {\bibfield  {journal} {\bibinfo  {journal}
  {Science}\ }\textbf {\bibinfo {volume} {343}},\ \bibinfo {pages} {157}
  (\bibinfo {year} {2014})}\BibitemShut {NoStop}%
\bibitem [{\citenamefont {Lundblad}\ \emph {et~al.}(2014)\citenamefont
  {Lundblad}, \citenamefont {Ansari}, \citenamefont {Guo},\ and\ \citenamefont
  {Moan}}]{Lundblad2014}%
  \BibitemOpen
  \bibfield  {author} {\bibinfo {author} {\bibfnamefont {N.}~\bibnamefont
  {Lundblad}}, \bibinfo {author} {\bibfnamefont {S.}~\bibnamefont {Ansari}},
  \bibinfo {author} {\bibfnamefont {Y.}~\bibnamefont {Guo}}, \ and\ \bibinfo
  {author} {\bibfnamefont {E.}~\bibnamefont {Moan}},\ }\href {\doibase
  10.1103/PhysRevA.90.053612} {\bibfield  {journal} {\bibinfo  {journal}
  {Physical Review A}\ }\textbf {\bibinfo {volume} {90}},\ \bibinfo {pages}
  {053612} (\bibinfo {year} {2014})}\BibitemShut {NoStop}%
\bibitem [{\citenamefont {Anderson}, \citenamefont {Kewming},\ and\
  \citenamefont {Turner}(2018)}]{Anderson2018}%
  \BibitemOpen
  \bibfield  {author} {\bibinfo {author} {\bibfnamefont {R.~P.}\ \bibnamefont
  {Anderson}}, \bibinfo {author} {\bibfnamefont {M.~J.}\ \bibnamefont
  {Kewming}}, \ and\ \bibinfo {author} {\bibfnamefont {L.~D.}\ \bibnamefont
  {Turner}},\ }\href {\doibase 10.1103/PhysRevA.97.013408} {\bibfield
  {journal} {\bibinfo  {journal} {Physical Review A}\ }\textbf {\bibinfo
  {volume} {97}},\ \bibinfo {pages} {013408} (\bibinfo {year}
  {2018})}\BibitemShut {NoStop}%
\bibitem [{\citenamefont {Trypogeorgos}\ \emph {et~al.}(2018)\citenamefont
  {Trypogeorgos}, \citenamefont {Vald{\'{e}}s-Curiel}, \citenamefont
  {Lundblad},\ and\ \citenamefont {Spielman}}]{Trypogeorgos2018}%
  \BibitemOpen
  \bibfield  {author} {\bibinfo {author} {\bibfnamefont {D.}~\bibnamefont
  {Trypogeorgos}}, \bibinfo {author} {\bibfnamefont {A.}~\bibnamefont
  {Vald{\'{e}}s-Curiel}}, \bibinfo {author} {\bibfnamefont {N.}~\bibnamefont
  {Lundblad}}, \ and\ \bibinfo {author} {\bibfnamefont {I.~B.}\ \bibnamefont
  {Spielman}},\ }\href {\doibase 10.1103/PhysRevA.97.013407} {\bibfield
  {journal} {\bibinfo  {journal} {Physical Review A}\ }\textbf {\bibinfo
  {volume} {97}},\ \bibinfo {pages} {013407} (\bibinfo {year}
  {2018})}\BibitemShut {NoStop}%
\bibitem [{\citenamefont {Davoodi}\ \emph {et~al.}(2019)\citenamefont
  {Davoodi}, \citenamefont {Jouda}, \citenamefont {Korvink}, \citenamefont
  {MacKinnon},\ and\ \citenamefont {Badilita}}]{Davoodi2019}%
  \BibitemOpen
  \bibfield  {author} {\bibinfo {author} {\bibfnamefont {H.}~\bibnamefont
  {Davoodi}}, \bibinfo {author} {\bibfnamefont {M.}~\bibnamefont {Jouda}},
  \bibinfo {author} {\bibfnamefont {J.~G.}\ \bibnamefont {Korvink}}, \bibinfo
  {author} {\bibfnamefont {N.}~\bibnamefont {MacKinnon}}, \ and\ \bibinfo
  {author} {\bibfnamefont {V.}~\bibnamefont {Badilita}},\ }\href {\doibase
  10.1016/j.pnmrs.2019.05.001} {\bibfield  {journal} {\bibinfo  {journal}
  {Progress in Nuclear Magnetic Resonance Spectroscopy}\ }\textbf {\bibinfo
  {volume} {112-113}},\ \bibinfo {pages} {34} (\bibinfo {year}
  {2019})}\BibitemShut {NoStop}%
\bibitem [{\citenamefont {Murphree}\ \emph {et~al.}(2007)\citenamefont
  {Murphree}, \citenamefont {Cahn}, \citenamefont {Rahmlow},\ and\
  \citenamefont {DeMille}}]{Murphree2007}%
  \BibitemOpen
  \bibfield  {author} {\bibinfo {author} {\bibfnamefont {D.}~\bibnamefont
  {Murphree}}, \bibinfo {author} {\bibfnamefont {S.~B.}\ \bibnamefont {Cahn}},
  \bibinfo {author} {\bibfnamefont {D.}~\bibnamefont {Rahmlow}}, \ and\
  \bibinfo {author} {\bibfnamefont {D.}~\bibnamefont {DeMille}},\ }\href
  {\doibase 10.1016/j.jmr.2007.05.025} {\bibfield  {journal} {\bibinfo
  {journal} {Journal of Magnetic Resonance}\ }\textbf {\bibinfo {volume}
  {188}},\ \bibinfo {pages} {160} (\bibinfo {year} {2007})}\BibitemShut
  {NoStop}%
\bibitem [{\citenamefont {Anders}\ and\ \citenamefont
  {Velders}(2018)}]{Anders2018}%
  \BibitemOpen
  \bibfield  {author} {\bibinfo {author} {\bibfnamefont {J.}~\bibnamefont
  {Anders}}\ and\ \bibinfo {author} {\bibfnamefont {A.~H.}\ \bibnamefont
  {Velders}},\ }in\ \href {\doibase 10.1002/9783527697281} {\emph {\bibinfo
  {booktitle} {Micro and Nano Scale NMR: Technologies and Systems}}},\ \bibinfo
  {series and number} {Advanced Micro and Nanosystems},\ \bibinfo {editor}
  {edited by\ \bibinfo {editor} {\bibfnamefont {J.}~\bibnamefont {Anders}}\
  and\ \bibinfo {editor} {\bibfnamefont {J.~G.}\ \bibnamefont {Korvink}}}\
  (\bibinfo  {publisher} {Wiley-VCH Verlag GmbH {\&} Co. KGaA},\ \bibinfo
  {address} {Weinheim, Germany},\ \bibinfo {year} {2018})\ p.\ \bibinfo {pages}
  {265}\BibitemShut {NoStop}%
\bibitem [{\citenamefont {Copeland}, \citenamefont {Robertson},\ and\
  \citenamefont {Verstraete}(1964)}]{Copeland1964}%
  \BibitemOpen
  \bibfield  {author} {\bibinfo {author} {\bibfnamefont {J.~R.}\ \bibnamefont
  {Copeland}}, \bibinfo {author} {\bibfnamefont {W.~J.}\ \bibnamefont
  {Robertson}}, \ and\ \bibinfo {author} {\bibfnamefont {R.~G.}\ \bibnamefont
  {Verstraete}},\ }\href {\doibase 10.1109/TAP.1964.1138196} {\bibfield
  {journal} {\bibinfo  {journal} {IEEE Transactions on Antennas and
  Propagation}\ }\textbf {\bibinfo {volume} {12}},\ \bibinfo {pages} {227}
  (\bibinfo {year} {1964})}\BibitemShut {NoStop}%
\bibitem [{\citenamefont {Kurpad}, \citenamefont {Wright},\ and\ \citenamefont
  {Boskamp}(2006)}]{Kurpad2006}%
  \BibitemOpen
  \bibfield  {author} {\bibinfo {author} {\bibfnamefont {K.~N.}\ \bibnamefont
  {Kurpad}}, \bibinfo {author} {\bibfnamefont {S.~M.}\ \bibnamefont {Wright}},
  \ and\ \bibinfo {author} {\bibfnamefont {E.~B.}\ \bibnamefont {Boskamp}},\
  }\href {\doibase 10.1002/cmr.b} {\bibfield  {journal} {\bibinfo  {journal}
  {Concepts in Magnetic Resonance Part B}\ }\textbf {\bibinfo {volume} {29B}},\
  \bibinfo {pages} {75} (\bibinfo {year} {2006})}\BibitemShut {NoStop}%
\bibitem [{\citenamefont {Heilman}\ \emph {et~al.}(2007)\citenamefont
  {Heilman}, \citenamefont {Riffe}, \citenamefont {Heid},\ and\ \citenamefont
  {Griswold}}]{Heilman2007}%
  \BibitemOpen
  \bibfield  {author} {\bibinfo {author} {\bibfnamefont {J.~A.}\ \bibnamefont
  {Heilman}}, \bibinfo {author} {\bibfnamefont {M.~J.}\ \bibnamefont {Riffe}},
  \bibinfo {author} {\bibfnamefont {O.}~\bibnamefont {Heid}}, \ and\ \bibinfo
  {author} {\bibfnamefont {M.~A.}\ \bibnamefont {Griswold}},\ }\href@noop {}
  {\bibfield  {journal} {\bibinfo  {journal} {Proceedings of the International
  Society of Magnetic Resonance in Medicine}\ }\textbf {\bibinfo {volume}
  {15}},\ \bibinfo {pages} {171} (\bibinfo {year} {2007})}\BibitemShut
  {NoStop}%
\bibitem [{\citenamefont {Lee}\ \emph {et~al.}(2009)\citenamefont {Lee},
  \citenamefont {Boskamp}, \citenamefont {Grist},\ and\ \citenamefont
  {Kurpad}}]{Lee2009}%
  \BibitemOpen
  \bibfield  {author} {\bibinfo {author} {\bibfnamefont {W.}~\bibnamefont
  {Lee}}, \bibinfo {author} {\bibfnamefont {E.}~\bibnamefont {Boskamp}},
  \bibinfo {author} {\bibfnamefont {T.}~\bibnamefont {Grist}}, \ and\ \bibinfo
  {author} {\bibfnamefont {K.}~\bibnamefont {Kurpad}},\ }\href {\doibase
  10.1002/mrm.21988} {\bibfield  {journal} {\bibinfo  {journal} {Magnetic
  Resonance in Medicine}\ }\textbf {\bibinfo {volume} {62}},\ \bibinfo {pages}
  {218} (\bibinfo {year} {2009})}\BibitemShut {NoStop}%
\bibitem [{\citenamefont {Gudino}\ \emph {et~al.}(2013)\citenamefont {Gudino},
  \citenamefont {Heilman}, \citenamefont {Riffe}, \citenamefont {Heid},
  \citenamefont {Vester},\ and\ \citenamefont {Griswold}}]{Gudino2013}%
  \BibitemOpen
  \bibfield  {author} {\bibinfo {author} {\bibfnamefont {N.}~\bibnamefont
  {Gudino}}, \bibinfo {author} {\bibfnamefont {J.~A.}\ \bibnamefont {Heilman}},
  \bibinfo {author} {\bibfnamefont {M.~J.}\ \bibnamefont {Riffe}}, \bibinfo
  {author} {\bibfnamefont {O.}~\bibnamefont {Heid}}, \bibinfo {author}
  {\bibfnamefont {M.}~\bibnamefont {Vester}}, \ and\ \bibinfo {author}
  {\bibfnamefont {M.~A.}\ \bibnamefont {Griswold}},\ }\href {\doibase
  10.1002/mrm.24462} {\bibfield  {journal} {\bibinfo  {journal} {Magnetic
  Resonance in Medicine}\ }\textbf {\bibinfo {volume} {70}},\ \bibinfo {pages}
  {276} (\bibinfo {year} {2013})}\BibitemShut {NoStop}%
\bibitem [{vcc()}]{vccsgit}%
  \BibitemOpen
  \href@noop {} {}\bibinfo {howpublished}
  {\url{https://github.com/JQIamo/RF-Current-Source}}\BibitemShut {NoStop}%
\bibitem [{\citenamefont {Karki}(1998)}]{Karki1998}%
  \BibitemOpen
  \bibfield  {author} {\bibinfo {author} {\bibfnamefont {J.}~\bibnamefont
  {Karki}},\ }\href@noop {} {\emph {\bibinfo {title} {{Voltage Feedback Vs
  Current Feedback Op Amps}}}},\ \bibinfo {organization} {Texas Instruments}
  (\bibinfo {year} {1998}),\ \bibinfo {note} {{SLVA051}}\BibitemShut {NoStop}%
\bibitem [{\citenamefont {Mancini}(2003)}]{Mancini2003}%
  \BibitemOpen
  \bibfield  {author} {\bibinfo {author} {\bibfnamefont {R.}~\bibnamefont
  {Mancini}},\ }\href {\doibase 10.1016/B978-0-7506-7701-1.X5000-7} {\emph
  {\bibinfo {title} {{Op Amps for Everyone}}}},\ edited by\ \bibinfo {editor}
  {\bibfnamefont {R.}~\bibnamefont {Mancini}}\ (\bibinfo  {publisher}
  {Elsevier},\ \bibinfo {year} {2003})\BibitemShut {NoStop}%
\bibitem [{ths(2015)}]{ths3091}%
  \BibitemOpen
  \href@noop {} {\emph {\bibinfo {title} {{THS309x High-voltage,
  Low-distortion, Current-feedback Operational Amplifiers}}}},\ \bibinfo
  {organization} {Texas Instruments} (\bibinfo {year} {2015}),\ \bibinfo {note}
  {{SLOS423H}}\BibitemShut {NoStop}%
\bibitem [{dis()}]{disclaimer}%
  \BibitemOpen
  \href@noop {} {}\bibinfo {note} {The identification of commercial products is
  for information only and does not imply recommendation or endorsement by the
  National Institute of Standards and Technology}\BibitemShut {NoStop}%
\bibitem [{psp(2016)}]{pspice}%
  \BibitemOpen
  \href@noop {} {\emph {\bibinfo {title} {{THS3091 PSpice Model}}}},\ \bibinfo
  {organization} {Texas Instruments} (\bibinfo {year} {2016}),\ \bibinfo {note}
  {{Rev. A}}\BibitemShut {NoStop}%
\bibitem [{\citenamefont {Balanis}(2016)}]{Balanis2016}%
  \BibitemOpen
  \bibfield  {author} {\bibinfo {author} {\bibfnamefont {C.~A.}\ \bibnamefont
  {Balanis}},\ }\href@noop {} {\emph {\bibinfo {title} {{Antenna Theory:
  Analysis and Design}}}},\ \bibinfo {edition} {4th}\ ed.\ (\bibinfo
  {publisher} {John Wiley and Sons, Inc.},\ \bibinfo {address} {Hoboken, NJ},\
  \bibinfo {year} {2016})\BibitemShut {NoStop}%
\bibitem [{Note1()}]{Note1}%
  \BibitemOpen
  \bibinfo {note} {The size of our antenna is set by geometric constraints of
  our laser-cooling apparatus.}\BibitemShut {Stop}%
\bibitem [{\citenamefont {Fratila}\ \emph {et~al.}(2014)\citenamefont
  {Fratila}, \citenamefont {Gomez}, \citenamefont {S{\'{y}}kora},\ and\
  \citenamefont {Velders}}]{Fratila2014}%
  \BibitemOpen
  \bibfield  {author} {\bibinfo {author} {\bibfnamefont {R.~M.}\ \bibnamefont
  {Fratila}}, \bibinfo {author} {\bibfnamefont {M.~V.}\ \bibnamefont {Gomez}},
  \bibinfo {author} {\bibfnamefont {S.}~\bibnamefont {S{\'{y}}kora}}, \ and\
  \bibinfo {author} {\bibfnamefont {A.~H.}\ \bibnamefont {Velders}},\ }\href
  {\doibase 10.1038/ncomms4025} {\bibfield  {journal} {\bibinfo  {journal}
  {Nature Communications}\ }\textbf {\bibinfo {volume} {5}},\ \bibinfo {pages}
  {3025} (\bibinfo {year} {2014})}\BibitemShut {NoStop}%
\bibitem [{ths(2018)}]{ths3491}%
  \BibitemOpen
  \href@noop {} {\emph {\bibinfo {title} {{THS3491 900-MHz, 500-mA High-Power
  Output Current Feedback Amplifier}}}},\ \bibinfo {organization} {Texas
  Instruments} (\bibinfo {year} {2018}),\ \bibinfo {note}
  {{SBOS875B}}\BibitemShut {NoStop}%
\bibitem [{\citenamefont {Eckel}\ \emph {et~al.}(2018)\citenamefont {Eckel},
  \citenamefont {Barker}, \citenamefont {Fedchak}, \citenamefont {Klimov},
  \citenamefont {Norrgard}, \citenamefont {Scherschligt}, \citenamefont
  {Makrides},\ and\ \citenamefont {Tiesinga}}]{Eckel2018}%
  \BibitemOpen
  \bibfield  {author} {\bibinfo {author} {\bibfnamefont {S.}~\bibnamefont
  {Eckel}}, \bibinfo {author} {\bibfnamefont {D.~S.}\ \bibnamefont {Barker}},
  \bibinfo {author} {\bibfnamefont {J.~A.}\ \bibnamefont {Fedchak}}, \bibinfo
  {author} {\bibfnamefont {N.~N.}\ \bibnamefont {Klimov}}, \bibinfo {author}
  {\bibfnamefont {E.}~\bibnamefont {Norrgard}}, \bibinfo {author}
  {\bibfnamefont {J.}~\bibnamefont {Scherschligt}}, \bibinfo {author}
  {\bibfnamefont {C.}~\bibnamefont {Makrides}}, \ and\ \bibinfo {author}
  {\bibfnamefont {E.}~\bibnamefont {Tiesinga}},\ }\href {\doibase
  10.1088/1681-7575/aadbe4} {\bibfield  {journal} {\bibinfo  {journal}
  {Metrologia}\ }\textbf {\bibinfo {volume} {55}},\ \bibinfo {pages} {S182}
  (\bibinfo {year} {2018})}\BibitemShut {NoStop}%
\bibitem [{doi()}]{doi}%
  \BibitemOpen
  \href@noop {} {}\bibinfo {howpublished}
  {\url{https://doi.org/10.5281/zenodo.3758489}}\BibitemShut {NoStop}%
\end{thebibliography}%


\begin{thebibliography}{4}%
\makeatletter
\providecommand \@ifxundefined [1]{%
 \@ifx{#1\undefined}
}%
\providecommand \@ifnum [1]{%
 \ifnum #1\expandafter \@firstoftwo
 \else \expandafter \@secondoftwo
 \fi
}%
\providecommand \@ifx [1]{%
 \ifx #1\expandafter \@firstoftwo
 \else \expandafter \@secondoftwo
 \fi
}%
\providecommand \natexlab [1]{#1}%
\providecommand \enquote  [1]{``#1''}%
\providecommand \bibnamefont  [1]{#1}%
\providecommand \bibfnamefont [1]{#1}%
\providecommand \citenamefont [1]{#1}%
\providecommand \href@noop [0]{\@secondoftwo}%
\providecommand \href [0]{\begingroup \@sanitize@url \@href}%
\providecommand \@href[1]{\@@startlink{#1}\@@href}%
\providecommand \@@href[1]{\endgroup#1\@@endlink}%
\providecommand \@sanitize@url [0]{\catcode `\\12\catcode `\$12\catcode
  `\&12\catcode `\#12\catcode `\^12\catcode `\_12\catcode `\%12\relax}%
\providecommand \@@startlink[1]{}%
\providecommand \@@endlink[0]{}%
\providecommand \url  [0]{\begingroup\@sanitize@url \@url }%
\providecommand \@url [1]{\endgroup\@href {#1}{\urlprefix }}%
\providecommand \urlprefix  [0]{URL }%
\providecommand \Eprint [0]{\href }%
\providecommand \doibase [0]{http://dx.doi.org/}%
\providecommand \selectlanguage [0]{\@gobble}%
\providecommand \bibinfo  [0]{\@secondoftwo}%
\providecommand \bibfield  [0]{\@secondoftwo}%
\providecommand \translation [1]{[#1]}%
\providecommand \BibitemOpen [0]{}%
\providecommand \bibitemStop [0]{}%
\providecommand \bibitemNoStop [0]{.\EOS\space}%
\providecommand \EOS [0]{\spacefactor3000\relax}%
\providecommand \BibitemShut  [1]{\csname bibitem#1\endcsname}%
\let\auto@bib@innerbib\@empty
\bibitem [{\citenamefont {Karki}(1998)}]{Karki1998}%
  \BibitemOpen
  \bibfield  {author} {\bibinfo {author} {\bibfnamefont {J.}~\bibnamefont
  {Karki}},\ }\href@noop {} {\emph {\bibinfo {title} {{Voltage Feedback Vs
  Current Feedback Op Amps}}}},\ \bibinfo {organization} {Texas Instruments}
  (\bibinfo {year} {1998}),\ \bibinfo {note} {{SLVA051}}\BibitemShut {NoStop}%
\bibitem [{ths(2015)}]{ths3091}%
  \BibitemOpen
  \href@noop {} {\emph {\bibinfo {title} {{THS309x High-voltage,
  Low-distortion, Current-feedback Operational Amplifiers}}}},\ \bibinfo
  {organization} {Texas Instruments} (\bibinfo {year} {2015}),\ \bibinfo {note}
  {{SLOS423H}}\BibitemShut {NoStop}%
\bibitem [{\citenamefont {Fratila}\ \emph {et~al.}(2014)\citenamefont
  {Fratila}, \citenamefont {Gomez}, \citenamefont {S{\'{y}}kora},\ and\
  \citenamefont {Velders}}]{Fratila2014}%
  \BibitemOpen
  \bibfield  {author} {\bibinfo {author} {\bibfnamefont {R.~M.}\ \bibnamefont
  {Fratila}}, \bibinfo {author} {\bibfnamefont {M.~V.}\ \bibnamefont {Gomez}},
  \bibinfo {author} {\bibfnamefont {S.}~\bibnamefont {S{\'{y}}kora}}, \ and\
  \bibinfo {author} {\bibfnamefont {A.~H.}\ \bibnamefont {Velders}},\ }\href
  {\doibase 10.1038/ncomms4025} {\bibfield  {journal} {\bibinfo  {journal}
  {Nature Communications}\ }\textbf {\bibinfo {volume} {5}},\ \bibinfo {pages}
  {3025} (\bibinfo {year} {2014})}\BibitemShut {NoStop}%
\bibitem [{psp(2016)}]{pspice}%
  \BibitemOpen
  \href@noop {} {\emph {\bibinfo {title} {{THS3091 PSpice Model}}}},\ \bibinfo
  {organization} {Texas Instruments} (\bibinfo {year} {2016}),\ \bibinfo {note}
  {{Rev. A}}\BibitemShut {NoStop}%
\end{thebibliography}%

\end{document}


\title{Supplemental Material for: A radiofrequency voltage-controlled current source for quantum spin manipulation}

\author{D.S. Barker}
\author{A. Restelli}
\affiliation{Joint Quantum Institute, University of Maryland and National Institute of Standards and Technology \\ College Park, MD 20742, USA}
\author{J.A. Fedchak}
\author{J. Scherschligt}
\author{S. Eckel}
\affiliation{Sensor Science Division, National Institute of Standards and Technology \\ Gaithersburg, MD 20899, USA}

\date{\today}

\maketitle

\textbf{Closed-loop transfer function:}
Applying Kirchoff's rules to the RF VCCS circuit (see Fig.~1 in the main text) and modelling the current-feedback amplifier (CFA) as suggested in Karki~\cite{Karki1998} yields:

\begin{widetext}
  \begin{equation}
    \label{eq:Kirchoff}
    \begin{split}
    V_{\text{out}} - V_s & = I_c L_c s,\\
    V_{\text{out}} - V_s & = \frac{I_p}{C_p s}, \\
    V_s & = I_s R_s, \\
    V_s - V_i & = I_f R_f, \\
    V_i & = I_g (R_g + L_g s), \\
    V_{\text{in}} - V_i & = R_i V_{\text{out}} \frac{1 + R_t C_t s}{R_t}, \\
    I_c + I_p & = I_f + I_s, \\
    I_g & = V_{\text{out}} \frac{1 + R_t C_t s}{R_t} + I_f. \\
    \end{split}
  \end{equation}
\end{widetext}

Here, \(s\) is the complex frequency; \(V_{\text{out}}\) is the output voltage of the CFA; \(V_s\) is the voltage across \(R_s\); \(V_i\) is the voltage at the inverting input of the CFA; \(V_{\text{in}}\) is the voltage at the noninverting input of the CFA; \(I_c\) is the current passing through the antenna coil \(L_c\); \(I_p\) is the current passing through \(C_p\); \(I_s\) is the current passing through \(R_s\); \(I_f\) is the current passing through \(R_f\); \(I_g\) is the current passing through \(R_g\); and the other symbols are defined in the main text.
We have also assumed that the antenna's parasitic resistance \(R_p\) is negligible.
Eliminating \(V_{\text{out}}\), \(V_i\), \(I_c\), \(I_p\), \(I_s\), \(I_f\), and \(I_g\) in Eqs.~\ref{eq:Kirchoff} leads to the closed-loop transfer function

\begin{widetext}
  \begin{equation}
    \label{eq:Closed}
    \begin{split}
      H(s) =\:&\frac{V_s}{V_{\text{in}}} \\
      =\:&\bigg[R_s \Big( R_f \big(R_t + R_t C_p L_c s^{2}\big) + \big(R_g + L_g s\big)\big(R_t + L_c s + R_t (C_p + C_t) L_c s^{2}\big)\Big)\bigg] \\
      &\Bigg/\,\bigg[R_f \Big(R_g + R_i + L_g s\Big)\Big(1 + R_t C_t s\Big)\Big(R_s + L_c s + R_s C_p L_c s^{2}\Big) \\
      &\; + s \Big(R_s L_c L_g s \big(1 + R_t (C_p + C_t) s\big) + R_s L_g \big(R_i + R_t + R_t R_i C_t s\big) \\
      &\; + L_c R_i \big(1 + R_t C_t s\big)\big(R_s + L_g s + R_s C_p L_g s^{2}\big)\Big) \\
      &\; + R_g \Big(R_i \big(1 + R_t C_t s\big)\big(R_s + L_c s + R_s C_p L_c s^{2}\big) + R_s \big(R_t + L_c s + R_t (C_p + C_t) L_c s^{2}\big)\Big)\bigg].
    \end{split}
  \end{equation}
\end{widetext}

The transadmittance of the RF VCCS is then
\begin{equation}
  Y(\omega) = \frac{\big|H(s\rightarrow j\omega)\big|}{R_s}.
\end{equation}
The linear response of the RF VCCS is given by \(Y(\omega)V_p\), where \(V_p\) is the RF input voltage amplitude.
We have found that the full transfer function of Eq.~\ref{eq:Closed} does not usefully guide selection of \(R_f\), \(R_g\), and \(L_g\).
Rather, we select \(R_f\) so that \(R_f + R_p + j\omega L_c\) approaches the recommended value of the feedback impedance as \(\omega\) approaches the operating bandwidth of the CFA and choose \(R_g\) to achieve the desired DC gain \(G\).~\cite{ths3091}
We then take \(C_p = 0\) and \(L_g = 0\) to reduce the number of poles in \(H(s)\), allowing us to find that the natural frequency \(\omega_n\) and damping factor \(\zeta\) are
\begin{equation}
  \begin{split}
  \omega_n = &\, \sqrt{\frac{1}{L_c C_t}}\sqrt{\frac{1+\frac{R_f}{R_t}+\frac{R_i}{R_g}\frac{R_f+R_g}{R_t}}{1+\frac{R_f}{R_s}+\frac{R_i}{R_g}\frac{R_s+R_f+R_g}{R_s}}}, \\
  \zeta = &\; \frac{1}{2}\sqrt{\frac{L_c}{R_t^{2}C_t}}\sqrt{\frac{R_t(R_f+R_s)}{R_s(R_f+R_t)}} \\
  + &\; \frac{1}{2}\sqrt{\frac{R_f^{2} C_t}{L_c}}\sqrt{\frac{R_t R_s}{(R_f+R_t)(R_f+R_s)}}.
  \end{split}
\end{equation}
Antennas with negligible \(R_p\) and \(L_c \gtrsim 10~\si{\nano\henry}\) will exhibit underdamped poles (\(\zeta < 1\)).
To flatten the response of the RF VCCS, we add a real pole to the transfer function using \(L_g\).
When \(R_i \ll R_g\), the pole introduced by \(L_g\) occurs at \(s=R_g/L_g\), so \(L_g\) must be larger than \(R_g/\omega_n\) to flatten \(Y(\omega)\).

\textbf{Open-loop transfer function:}
To find the open-loop transfer function, we ground the noninverting input of the CFA and break the feedback loop at the CFA's output.
The input voltage \(V_{\text{in}}\) is now applied directly to the antenna.
Applying Kirchoff's rules to the open-loop circuit leads to
\begin{equation}
  \label{eq:Kirchoff2}
  \begin{split}
  V_{\text{in}} - V_s & = I_c L_c s,\\
  V_{\text{in}} - V_s & = \frac{I_p}{C_p s}, \\
  V_s & = I_s R_s, \\
  V_s - V_i & = I_f R_f, \\
  V_i & = I_g (R_g + L_g s), \\
  V_i & = -I_e R_i, \\
  V_{\text{out}} & = \frac{I_e R_t}{1 + R_t C_t s}, \\
  I_c + I_p & = I_f + I_s, \\
  I_g & = I_e + I_f, \\
  \end{split}
\end{equation}
where \(I_e\) is the error current.~\cite{Karki1998}
We eliminate \(V_s\), \(V_i\), \(I_c\), \(I_p\), \(I_s\), \(I_f\), \(I_g\), and \(I_e\) in Eqs.~\ref{eq:Kirchoff2} to find the open-loop transfer function

\begin{widetext}
  \begin{equation}
    \begin{split}
      G(s) =\:&\frac{V_{\text{out}}}{V_{\text{in}}} \\
      =\:&-\bigg[R_s R_t \big(R_g + L_g s\big)\big(1 + C_p L_c s^2\big)\bigg] \\
      &\Bigg/\,\bigg[\Big(1 + C_t R_t s\Big)\Big(R_g \big(R_i R_s + L_c (R_i + R_s) s + C_p L_c R_i R_s s^{2}\big) \\
      &\; + R_f \big(R_g + R_i + L_g s\big)\big(R_s + L_c s + C_p L_c R_s s^2\big) \\
      &\; + s \big((L_c + L_g) R_i R_s + L_c L_g (R_i + R_s) s + C_p L_c L_g R_i R_s s^2\big)\Big)\bigg].
    \end{split}
  \end{equation}
\end{widetext}

Once we select component values using \(H(s)\) (see above), we can compute \(G(s)\) to estimate the gain margin and phase margin of the RF VCCS.\@

\begin{figure}
  \includegraphics[width=4.5in]{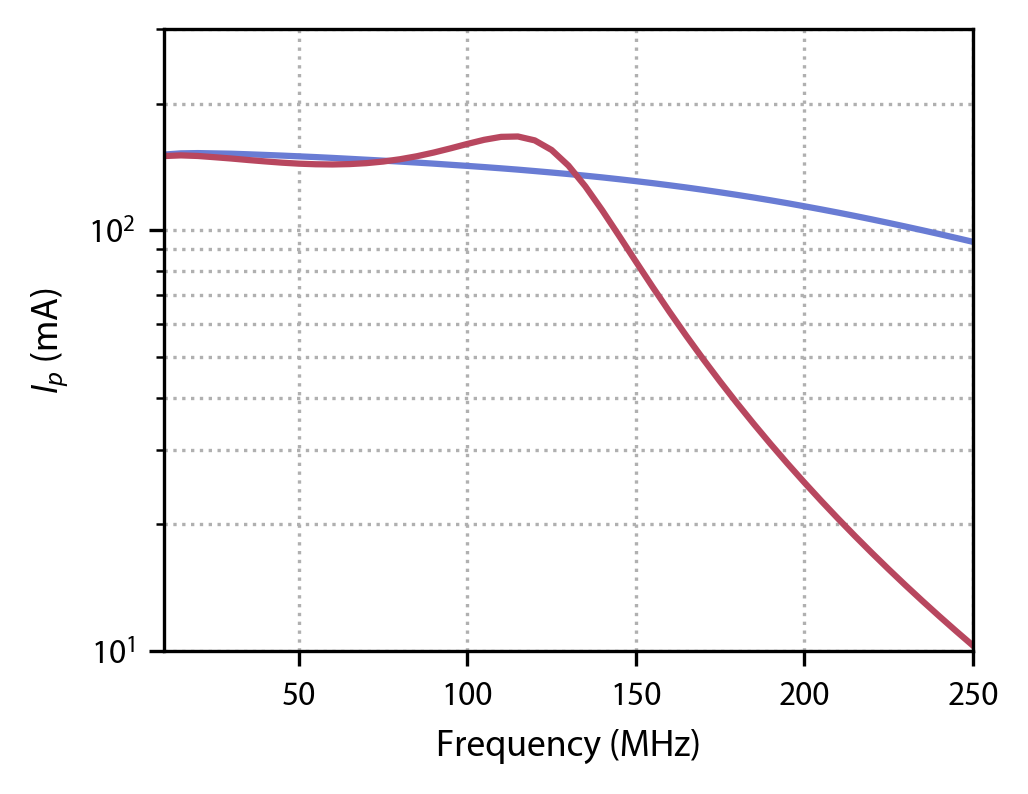}
  \caption{\label{Fig:Antennas}
  Simulations of a \(25~\si{\nano\henry}\) antenna (red) and the design from the supplemental material of Fratila et al.~\cite{Fratila2014} (blue).
  The RF input voltage for both curves is \(V_p = 127~\si{\milli\volt}\).
  }
\end{figure}

\textbf{Low-inductance antennas:}
We have simulated the performance of two low-inductance antennas.
The first is an \(L_c = 25~\si{\nano\henry}\) loop antenna with \(R_p = 100~\si{\milli\ohm}\) and \(C_p = 1~\si{\pico\farad}\).
The second is the broadband coil design described in the supplemental material of Fratila et al.~\cite{Fratila2014} with \(L_c = 0.4~\si{\nano\henry}\), \(R_p = 5~\si{\ohm}\), and \(C_p = 0~\si{\pico\farad}\).
The results of transient simulations of the two antennas are shown in Fig.~\ref{Fig:Antennas}.
For the \(25~\si{\nano\henry}\) antenna simulation, \(R_f = 561~\si{\ohm}\), \(R_g = 62~\si{\ohm}\), \(R_s = 8.33~\si{\ohm}\) (after accounting for the parallel impedance of a hypothetical spectrum analyzer), \(L_g = 130~\si{\nano\henry} \approx 1.3 R_g/\omega_n\).
For the broadband coil simulation, \(R_f = 866~\si{\ohm}\), \(R_g = 95.3~\si{\ohm}\), \(R_s = 8.33~\si{\ohm}\), \(L_g = 0~\si{\nano\henry}\).
The simulated bandwidth of the RF VCCS is approximately \(140~\si{\mega\hertz}\) for the \(25~\si{\nano\henry}\) antenna and approximately \(210~\si{\mega\hertz}\) for the broadband coil.
We note that the THS3091 SPICE model underestimates the device bandwidth for \(G = 10\),~\cite{pspice} so wider operating bandwidths may be achievable in practice.

\bibliography{RFVI}